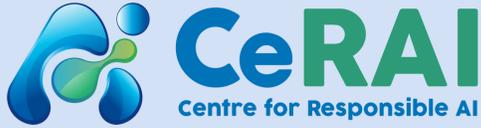

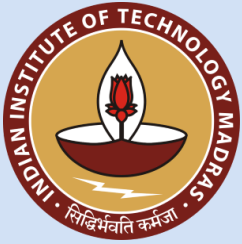

# Participatory Approaches in AI Development and Governance | Case Studies

**Centre for Responsible AI**

Robert Bosch Centre for Data Science and AI,
IIT Madras,
Chennai 600036

contact@cerai.in

twitter.com/cerai iitm
linkedin.com/company/cerai- iitm

In collaboration with

**V | D H | Centre for Legal Policy**

# About us

## Centre for Responsible AI

Centre for Responsible AI ('CeRAI') is a virtual interdisciplinary research centre at IIT Madras focused on fundamental and applied research in Responsible AI. Our research includes technical research on making AI models and products fair, understandable and safe, policy analysis on governance and regulation of AI, and substantial training and outreach programmes addressing a wide audience ranging from primary school children to senior executives.

Website: https://cerai.iitm.ac.in/

## Vidhi Centre for Legal Policy

Vidhi Centre for Legal Policy ('Vidhi') is an independent think-tank undertaking legal research to make better laws and improve governance for the public good. We do this through high quality, peer reviewed original legal research, engaging with the Government of India, State governments and other public institutions to both inform policy making and to effectively convert policy into law. Since 2013, Vidhi has worked with over 20 different ministries in the Government of India, 9 State Governments, the Supreme Court of India, 14 regulators and public institutions for different projects, including 77 enactments of binding law or policy. During this period, Vidhi has also produced over 369 pieces of original legal research.

Website: www.vidhilegalpolicy.in

# About the Authors

## CeRAI

**Ambreesh Parthasarathy** is a Post Baccalaureate Fellow at the Centre for Responsible AI (CeRAI) at IITM.

**Prof. Balaraman Ravindran** is the head of the Centre for Responsible AI (CeRAI) at IITM.

**Gokul Krishnan** is a Research Scientist at the Centre for Responsible AI (CeRAI) at IITM.

## Vidhi

**Ameen Jauhar** was a Senior Resident Fellow and Team Lead at the Centre for Applied Law and Technology (ALTR) at Vidhi.

**Aditya Phalnikar** is a Research Fellow at the Research Director's Office at Vidhi.

The authors would also like to thank Shehnaz Ahmed, Sudarsan Padmanabhan, Shibu Guruvayur, Sanjay Karanth and Jyotsana Singh for their comments and feedback.

Errors in the paper, if any, are the sole responsibility of the authors.

# Table of Contents

## *Introduction*

This paper forms the second of a two-part series on the value of a participatory approach to AI development and deployment. The first paper had crafted a principled, as well as pragmatic, justification for deploying participatory methods in these two exercises (that is, development and deployment of AI). The pragmatic justification is that it improves the quality of the overall algorithm by providing more granular and minute information. The more principled justification is that it offers a voice to those who are going to be affected by the deployment of the algorithm, and through engagement attempts to build trust and buy-in for an AI system. By a participatory approach, we mean including various stakeholders (defined a certain way) in the actual decision making process through the life cycle of an AI system.

Despite the justifications offered above, actual implementation depends crucially on how stakeholders in the entire process are identified, what information is elicited from them, and how it is incorporated. **This paper will test these preliminary conclusions in two sectors – the use of [facial recognition technology](#) in the upkeep of [law and order](#) and the use of [large language models in the healthcare sector](#).** These sectors have been chosen for two primary reasons. Since Facial Recognition Technologies ('FRTs') are a branch of AI solutions that are well-researched and the impact of which is well documented, it provides an established space to illustrate the various aspects of adapting PAI to an existing domain, especially one that has been quite contentious in the recent past. LLMs in healthcare provide a canvas for a relatively less explored space, and helps us illustrate how one could possibly envision enshrining the principles of PAI for a relatively new technology, in a space where innovation must always align with patient welfare.

As we will see, the deployment of AI is not an uncontested good. While it can deliver benefits in the way of increased efficiency, some matters must first be accounted for. The data on which the algorithm is trained needs to be representative of the population upon which it is later deployed. It needs to be ensured that the scope for bias is reduced to a minimum. Even before the question of representativeness can be decided, there ought to be a general discussion on whether the AI solution is actually required in that context, and an ex-ante impact assessment should be conducted. This paper adds on to paper 1 by ensuring that there is a more contextualised discussion on what problems can arise in the implementation of AI softwares, and how they can be dealt with. It should hence be seen as a comprehensive illustration of the principles enunciated in paper 1.

The paper is divided into two sections with similar structures. The first section deals with the use of facial recognition technology in the upkeep of law and order, and the second deals with the use of LLMs in the healthcare sector. For each section, a brief introduction of the technology is first given. Following this, the scope of the technology in the sector concerned is discussed, after which issues with this implementation are enumerated. Finally, we utilise the decision sieve (outlined in Paper 1) to discuss how, in theory, a participatory approach can help mitigate some risks of potential harms associated with each technology. A brief foray into what this means in the Indian regulatory context follows.

# FRT in Law Enforcement (an in-depth view)

Facial Recognition Technology ('FRT') can be traced back as far as the 1960s[1]. These techniques were based around hand coding facial features for every individual face in the database. The rise of databases and readily available image data, automation, machine learning etc. have ushered in a new era for this technology. Facial recognition has become a mainstay in the tech fold and is used in various contexts such as unlocking one's phone, clinical diagnostics, manufacturing etc.

FRT is typically used for two key purposes: verification and identification[2]. Verification is done by matching the live photograph of a person to the pre-existing photograph that is on the authority's database (1:1). Use of FRT for verification is done to authenticate the identity of an individual seeking to gain access to any benefits or government schemes. Identification is done by trying to get a match between the face of an individual which has been extracted from a photograph/video and the entire database of the authority in order to ascertain the identity of the individual (1:many). Use of FRT for identification is usually done for the purposes of security and surveillance[3]. The only official publication on FRTs as of the time of writing this paper is NITI Aayog's Responsible AI for All discussion paper, which takes FRTs as one of its use cases for its framework[4].

As noted earlier, one of the most well-known approaches of facial recognition is its use in law enforcement. Agencies often use it to monitor crowds, perform general vigilance, and find people (missing or otherwise). It was noted that 42 US Federal Agencies that employ law enforcement officers have used the technology in one form or the other[5]. It was also found by the Internet Freedom Foundation ('IFF'), through an RTI, that the Delhi Police had been using FRTs for surveillance and policing. With an arbitrary mark of 80% to be called a positive match, the tech which was originally used to identify missing children (*Sadhan Haldar v NCT*)[6] was used for policing, with no rigorous guardrails against misidentification or false positives.

---

[1] A Facial Recognition Project Report, (Woodrow Wilson Bledsoe) <https://archive.org/details/firstfacialrecognitionresearch/FirstReport/>.
[2] It is to be noted that these are not the only use cases of FRTs. Surveillance and monitoring are well documented use cases by law enforcement.
[3] From investigation to conviction: How does the Police use FRT?, (Internet Freedom Foundation); <https://internetfreedom.in/from-investigation-to-conviction-how-does-the-police-use-frt/#:~:text=So%20how%20is%20FRT%20evidence,build%20the%20case%20against%20them>.
[4] Responsible AI for All Discussion Paper, (NITI Aayog), <https://www.niti.gov.in/sites/default/files/2022-11/Ai_for_All_2022_02112022_0.pdf>.
[5] Facial Recognition Technology: Federal Law Enforcement Agencies Should Have Better Awareness of Systems Used By Employees, (U.S. Government Accountability Office); <https://www.gao.gov/products/gao-21-105309>.

In fact, the design, development, and deployment of FRT systems have been embraced by the state in its public functions and by private actors. For example, FRTs are being used for law enforcement in Tamil Nadu (FaceTagr), Punjab (PAIS), Uttar Pradesh (Trinetra) and New Delhi (AI Vision).[7] Additionally, FRT is being used for delivery of services and increasing efficiency in governance such as the Real Time Digital Authentication of Identity project to authenticate pensioners in Telangana[8], the DigiYatra authentication project being carried out in select airports across India[9], and the Face Matching Technology system adopted by the Central Board for Secondary Education to provide access to academic documents by authenticating a student's identity through FRT.[10]

---

[6] W.P. (Crl.) 1560/2017. The case is accessible at <https://drive.google.com/file/d/1toYSlQpnveVpXuHMmPGqoSjVP7EcY3jU/view?ref=static.internetfreedom.in>.

[7] Divya Chandrababu, 'Facial recognition system of Tamil Nadu police stirs privacy row' (*Hindustan Times* December 10 2022) <https://www.hindustantimes.com/india-news/facial-recognition-system-of-tn-police-stirs-privacy-row-101670614143523.html> accessed 28th September 2023; Gopal Sathe, 'Cops In India Are Using Artificial Intelligence That Can Identify You In a Crowd' (*Huffpost* August 16 2018) <https://www.huffpost.com/archive/in/entry/facial-recognition-ai-is-shaking-up-criminals-in-punjab-but-should-you-worry-too_in_5c107639e4b0a9576b52833b> accessed 28th September 2023; 'Staqu launches TRINETRA, an AI app for UP Police Department' (*Deccan Chronicle* December 29 2018) <https://www.deccanchronicle.com/technology/in-other-news/291218/staqu-launches-trinetra-an-ai-app-for-up-police-department.html> accessed 28th September 2023; Varsha Bansal, 'The Low Threshold for Face Recognition in New Delhi' (*Wired* 21 August 2022) <https://www.wired.co.uk/article/delhi-police-facial-recognition> accessed 28th September 2023.

[8] 'Telangana government leveraging the power of AI and ML for pensioners' (*IndiaAI* October 27 2022) <https://indiaai.gov.in/case-study/telangana-government-leveraging-the-power-of-ai-and-ml-for-pensioners> accessed 28th Septemeber 2023.

[9] Responsible AI for All Discussion Paper, (NITI Aayog), <https://www.niti.gov.in/sites/default/files/2022-11/Ai_for_All_2022_02112022_0.pdf>; Saurabh Sinha, 'DigiYatra Roll out: Your face will now be an ID & domestic boarding card at Delhi, Bengaluru and Varanasi airports' (*Times of India* December 5 2022). <https://timesofindia.indiatimes.com/business/india-business/digiyatra-rolled-out-your-face-will-now-be-your-id-and-domestic-ticket-at-delhi-bengaluru-and-varanasi/articleshow/95912778.cms> accessed 28th September 2023.

[10] 'CBSE introduces 'Facial Recognition System' for accessing digital academic documents of Class 10 and 12' (*Hindustan Times* October 22 2020) <https://www.hindustantimes.com/education/cbse-introduces-facial-recognition-system-for-accessing-digital-academic-documents-of-class-10-and-12/story-XmoqbgNRCeVD9X91zzxFzM.html> accessed 28th September 2023.

The way FRT is implemented is novel. With a large corpus of images, it seeks to extract meaningful features of the face and carry out predictions. While this might cause excitement to some, as it is viewed by some as a technological aid that would improve quality of policing, many raise various concerns. Two of the major concerns raised are the issues of *privacy* and *accuracy*. Several important and pertinent questions like how data is procured, what the demographic in the training data looks like, whether the extracted features are relevant, what the accuracy is and how different demographic subgroups affect the accuracy, should be asked and addressed to understand the impact of the technology. While many agencies like Interpol and WEF have come together to list out best practices[11&12], there are fundamental questions that must be looked at while implementing the technology. The following sections attempt to delve into the various aspects of FRTs, their implementation, the possible risks and how we could possibly mitigate them.

## How FRTs work

Computer Vision algorithms "see" images and analyse them in a manner quite different to humans. Our ability to see is primarily powered by our ability to perceive and identify different features in the image from prior knowledge. We are able to identify, group, and separate these elements and thus make sense of relevant information. Thus, we are able to identify and recognise faces and other objects in a given image.

In contrast to this, a computer vision algorithm only sees a grid of pixel values. These are matrices (rows and columns) of numbers indicating the nature of colour, its intensity, its brightness etc. In order to detect an object, or in this case a face, a computer vision algorithm must take this array of numbers, and point out patterns in the numerical values that reliably reflect facial features. There have been various approaches taken over the years to tackle this problem statement of reliably identifying these patterns that reflect facial features. Early techniques focused on a top-down approach. They resorted to encoding human knowledge of facial features and applied this encoding to the grid to find the corresponding patterns. This approach had two major issues. First, the imprecision of these fairly generic rules could result in multiple faces that reflect the same rule. Second, the images would need to be front-facing, with environmental factors like lighting and occlusions and elements like facial orientation and expression kept relatively uniform. These issues spurred the search for alternative methods to solve the problem statement. This led to a shift to a bottom-up approach. In 2001, researchers Paul Viola and Michael Jones[13] came up with a path-breaking algorithm for detecting faces. Instead of directly trying to extract facial features, they scanned through images using rectangular frames known as Haar features to detect common patterns that allowed for rapid facial detection. These features were edges, lines, and diagonals of different scales. They trained this algorithm on a dataset of face images and non-face images to tighten the boundary of these rectangular frames, narrowing down their space to the most important ones. To detect a face, quantified, pixel value expressions of each of these features would be placed over the image's numerical grid and slid across the entire picture from box to box in order to find those areas where matching changes in brightness intensity uncovered corresponding matches with facial patterns in the image.

---

[11] The Presidio Recommendations on Responsible Generative AI, (World Economic Forum) <https://www3.weforum.org/docs/WEF_Presidio_Recommendations_on_Responsible_Generative_AI_2023.pdf>.
[12] Responsible AI Innovation in Law Enforcement, (Interpol) <https://unicri.it/sites/default/files/2024-02/00_README_File_Feb24.pdf>.

But this algorithm came with its own set of problems. Though it was good at identifying faces from non-faces, it could not distinguish between faces or pairs of the same face among many. But it provided the basis for modern FRT techniques by promoting a data-based approach that could aspire to extract deeper and more nuanced latent features. This would help the algorithm to tell us whether two photos represented the same face or if a face matched any others in a large pool.

The model that was up to this task was the Convolutional Neural Network (CNN)[14]. Unlike the Viola Jones algorithm, CNNs use kernels, a small grid of pixel values (a matrix of pixel values), that create feature maps through the entire image trying to identify matches for the particular feature that they have been trained to look for. A CNN thus, is a set of filters stacked over each other, where each filter identifies a particular pattern (like an edge or corner, or geometric features like circles and squares). As we dive deeper, the pattern that is identified becomes increasingly more sophisticated (ears, eyes, nose etc) while layers that are deeper are able to identify man, woman, dog etc. This can be viewed as the aggregation of simpler features to understand and identify more complex patterns, much like how to identify a wheel – you could first identify that it's a circle, then that it has spokes in it, and finally that it seems to look similar to other wheels we've seen. This sort of method of learning is completely supervised, thus learning features from the given data, instead of using handcrafted features. The initial kernels that are initialised to random values, back-propagate the error (difference between the pixel values in the kernel and the actual picture) and try to make adjustments to fit the curve of the different images and their pixel values. Thus, this method captures the nuance of the dataset instead of trying to fit into our rigid definitions of what features must be. However, there are still limits to what this method can achieve. Uncontrolled environments have shown to give accuracy of matches anywhere between 36% and 87% depending on the position of the cameras. For instance, it cannot accurately identify people wearing masks, which has become commonplace post COVID[15].

---

[13] Rapid Object Detection using a Boosted Cascade of Simple Features, <https://www.cs.cmu.edu/~efros/courses/LBMV07/Papers/viola-cvpr-01.pdf>.

[14] Deep Learning, LeCun, Yann & Bengio, Y. & Hinton, Geoffrey. (2015). Deep Learning. Nature. 521. 436-44. 10.1038/nature14539., <http://www.cs.toronto.edu/~hinton/absps/NatureDeepReview.pdf>.

One can also intuitively observe that this method has an inherent problem. This method's greatest strength – being driven by data – also highlights a possible vulnerability. It is highly dependent on the quality of data that is provided to train and might struggle to identify differences that are represented by different demographics. For example, physiological differences between able-bodied people, and differently abled people, differences between different races (like complexion and skin tones) with drastically different facial features would be hard for the model to identify, if it hasn't been trained on a sufficiently large amount of data representative of these features.

This also works the other way as applications like facial recognition in law enforcement, might have datasets that are institutionally skewed or biassed (through the biases in legacy datasets), leading to subpar performance over different demographics, and to discriminatory policing. Datasets that are imbalanced are not a new problem to AI and there have been many methods to deal with these issues. However, poor model design that does not acknowledge these imbalances and does not test rigorously for different demographics, especially in the context of FRT in law enforcement, exacerbates the problem of increased false positives or false negatives.

There are other issues that arise with the implementation of these techniques. FRTs being procured through the private sector is a massive cause for concern. With privacy being the overarching concern, there are also issues of delegation of surveillance to the public, transparency around decision making and private sector incentives driving public policy[16]. Disproportionate implementation of FRTs has also shown to lead to selective and disproportionate policing over historically disadvantaged minorities[17].

---

[15] Facial Recognition Technology in Law Enforcement in India, (IDFC Institute);, <https://www.idfcinstitute.org/site/assets/files/16530/facial_recognition_technology_in_law_enforcement_in_india.pdf>.

While FRTs, as a concept, hold some value and promise to ease the efficiency burden on law enforcement agencies, the technology at present is not mature enough to be completely relied upon[18]. This begs the question of how we can improve the implementation to ensure it is done in a safe manner. To do so, we must understand and acknowledge the issues and limitations of FRTs as an AI based solution.

## *Overview of FRT deployment in Indian policing*

Unfortunately, there is no official data available on how and for what purposes the Indian police forces are using these technologies.[19] From a review of the available literature, it emerges that FRT is being implemented for two (not mutually exclusive) purposes – monitoring and investigation[20&21]. Most recently, the Bangalore police has started their project 'Safe City'[22], which uses FRT to detect illegal parking, identify faces from a 'blacklist' generated by the city's police (although the method of generation is unclear).[23] As per the work order released by the Bengaluru police after an RTI was filed, it was also intended to be used for motion path analysis and identifying violent activities in a crowd[24]. Once again, the method by which it was being done was unclear.

---

[16] Procurement of Facial Recognition Technology for Law Enforcement in India: Legal and Social Implications of the Private Sector's Involvement, (Jauhar and Vipra, Vidhi Centre for Legal and Policy Design);
<https://vidhilegalpolicy.in/research/procurement-of-facial-recognition-technology-for-law-enforcement-in-india-legal-and-social-implications-of-the-private-sectors-involvement/>,
<https://vidhilegalpolicy.in/wp-content/uploads/2021/12/FRT-paper-3-Vidhi-format-2.pdf>.
[17] The Use of Facial Recognition Technology for Policing in Delhi, (Vipra, Vidhi Centre for Legal and Policy Design);
< https://vidhilegalpolicy.in/research/the-use-of-facial-recognition-technology-for-policing-in-delhi/>.
[18] ibid.
[19] For example, in a previous working paper, researchers at Vidhi have highlighted that in many instances, police forces are not responding to RTI requests filed about how exactly FRT technologies are being used and who they are being procured from.
[20] ibid.
[21] Indian Express, Face tech behind Delhi riots arrests: 'Accused told to match pose';
<https://indianexpress.com/article/cities/delhi/delhi-riots-arrests-accused-told-to-match-pose-8083399/>
[22] Panoptic Project, 'Bengaluru' accessed here.
[23] ibid.

In 2021, the Bihar police had floated a tender for an integrated surveillance system that could, among other things, match the faces of suspects to various databases.[25] FRT systems are also used widely in Telangana and Delhi.[26] In Hyderabad, the city's police commissioner has filed an affidavit before the Telangana High Court to the effect that the FRT tool is used to compare 'suspicious' persons against a database of offenders, missing persons, etc.[27] The affidavit also states that the CCTV system in place in the city is completely separate from the FRT system, hence ensuring that there is no mass surveillance. However, there is not enough clarity on this point, as reports by Amnesty state that the CCTV system can be integrated with the FRT system.[28]

The Delhi Police also uses FRT technology to match suspects against photographs in databases. As per an RTI reply filed in response to a query by the IFF, they use a match of 80% as threshold to be considered a positive match, and might investigate further in case the match is below 80%.[29] FRT technologies are also being used for conflict area monitoring by the Indian Army.[30]

---

[24] Available here https://drive.google.com/drive/folders/189C5-6wrT6KfIZu5zr2MFK6mDjl6JPht.
[25] Aihik Sur. Bihar Looking To Deploy Facial Recognition System In Bhagalpur And Muzaffarpur, Connect It To CCTNS. Medianama. April 19, 2021. Available at:
https://www.medianama.com/2021/04/223-bihar-bhagalpurmuzaffarpur-facial-recognition-cctns/.
[26] Anushka Jain, 'Delhi Police's claims that FRT is 80% accurate are 100% scary' (*IFF* August 17 2022) <https://internetfreedom.in/delhi-polices-frt-use-is-80-accurate-and-100-scary/> accessed 16th February 2024; Vandana Menon, 'Hyderabad wants to be smart, efficient. But face recognition tech, CCTVs making it paranoid' (*The Print* 22 December 2023) <https://theprint.in/ground-reports/hyderabad-wants-to-be-smart-efficient-but-face-recognition-tech-cctvs-making-it-paranoid/1895948/> accessed 16th February 2024.
[27] Balakrishna Ganeshan, 'No mass surveillance using facial recognition in Hyderabad, top cop tells HC' (*News Minute* January 8 2023) <https://www.thenewsminute.com/telangana/no-mass-surveillance-using-facial-recognition-hyderabad-top-cop-tells-hc-171703> accessed 16th February 2024.
[28] Amnesty International, 'India: Hyderabad 'on the brink of becoming a total surveillance city'' (*Amnesty International* November 9 2021) <https://www.amnesty.org/en/latest/news/2021/11/india-hyderabad-on-the-brink-of-becoming-a-total-surveillance-city/> accessed 16th February 2024.
[29] ibid.
[30] Huma Siddiqui, 'DRDO develops advanced Facial Recognition Technology (FRT) to boost surveillance' (*Financial Times* August 22 2022) <https://www.financialexpress.com/business/defence-drdo-develops-advanced-facial-recognition-technology-frt-to-boost-surveillance-2639708/> accessed 16th February 2024.

Overall, it becomes clear that police forces in India are beginning to use FRT on a systematic scale for surveillance, monitoring, and investigation. As we have seen, for most cities, these functions are also not exclusive to each other. Once an FRT AI is connected to the city's CCTV systems, it is possible to monitor individuals as part of a general law and order programme and also match the faces of suspects to defined databases.

## *Issues:*

### *Bias & Fairness*

Bias in computational systems is defined as a systematic error that results in unfair outcomes[31]. In the context of AI it can arise from various sources, some of them being data collection, algorithm design, human interpretation etc. Machine Learning models, trained on data biassed in some form are often found to replicate patterns of bias present in the training data. It is important that these biases are identified and worked on. The presence of these quantifiable biases has largely driven discourse on fairness in AI-based solutions. Fairness, usually looked at as the absence of bias, for example algorithmic bias, has been found to be a very important lens to judge the efficacy of AI systems. In our case, we would be taking a slightly more nuanced approach to fairness, as in most cases algorithmic biases enable model performance. It is to be noted that algorithmic biases are different from societal biases. Algorithmic biases pertain to abstract patterns that are learnt during training that help do the task better. The issues arise when some of these learnt patterns reinforce societal biases.Fairness as used here has to be understood as the presence of algorithmic or data bias that is resulting in societal bias. This will help us understand the real world implications and performance much better.

In the case of Facial Recognition Technology, there are various biases that need to be acknowledged. Firstly, the bias that comes with the capture of images themselves[32]. It has been very well documented that image-capture technology, from the days of film, has had a history of occurrences where the technology is tuned to capture certain colours and profiles better. When the actual contrast and colour gradients are represented poorly in the pixel array, the training of the model and the subsequent inferences would also carry the same biases. Another source of bias is the algorithmic bias, where the algorithm hasn't accounted for subpar performance for different sub-groups in the dataset. If the set of features the algorithm uses does not account for the differences among these various sub-groups, then it is bound to perform sub-optimally owing to the bias.

---

[31] Ferrara, Emilio, Fairness and Bias in Artificial Intelligence: A Brief Survey of Sources, Impacts, and Mitigation Strategies (October 27, 2023). Available at SSRN: <https://ssrn.com/abstract=4615421> or <http://dx.doi.org/10.2139/ssrn.4615421>.
[32] Ruha Benjamin, Race after Technology: Abolitionist Tools for the New Jim Code (Cambridge, UK: Polity

As mentioned earlier, the learning setting is primarily supervised. Not having the right labels to annotate sub-groups leads to suboptimal learning results. For example, UTKFace that was published in 2017 had only labels for 5 groups, which were White, Black, Asian, Indian and Others. This shows us that for downstream applications like law enforcement, where the margin for error is very minimal, such datasets do not do justice to the representation required to get the outcomes from training. This is especially true in the Indian context. With the diversity in physiology over the various different geographical regions, datasets that do not represent sub-demographics well, and algorithms that don't account for this diversity, will lead to repercussions in downstream tasks. This is extremely concerning given that the use of FRTs in law enforcement has been exponentially trending upwards recently.[33] The use of FRT is taking place in India without there being any standards in place to regulate the technology or certify its quality. As an example, the Bengaluru police RTI response states this quite explicitly.[34] Thus, there is a very real possibility that a sub-par FRT system is adopted by the Police which leads to misidentification, and ultimately, a false

---

Press, 2019).; Simone Browne, Dark Matters: on the Surveillance of Blackness (Durham, NC: Duke University Press, 2015).; Dyer, White.; Sarah Lewis, "The Racial Bias Built into Photography," The New York Times, April 25, 2019.; Lorna Roth, "Looking at Shirley, the Ultimate Norm: Colour Balance, Image Technologies, and Cognitive Equity," Canadian Journal of Communication 34, no. 1 (2009), https://doi.org/10.22230/cjc.2009v34n1a2196 ; Lorna Roth, "Making Skin Visible through Liberatory Design," in CaptivatingTechnology: Race, Carceral Technoscience, and Liberatory Imagination in Everyday Life, ed. Ruha Benjamin (Durham, NC: Duke University Press, 2019).

[33] Facing Up To the Risks of Automated Facial-Recognition Technologies in Indian Law Enforcement, (Jauhar, Vidhi Centre for Legal and Policy Design); <https://vidhilegalpolicy.in/research/facing-up-to-the-risks-of-automated-facial-recognition-technologies-in-indian-law-enforcement/>.

[34] Meet the Facial Recognition Giant Helping Bengaluru Police <https://analyticsindiamag.com/meet-the-facial-recognition-giant-helping-bangalore-police/#:~:text=The%20RTI%20responses%20received%20by,detects%20a%20person%20on%20the>.

13conviction. It was found as a response to RTIs filed by the Internet Freedom Foundation[35] that the Delhi Police was using the mark of 80% as the threshold to decide whether a match was positive or a false positive. While there might have been multiple rounds of discussions to decide this threshold value, it still does not mitigate the issue of not having a body that regulates the standard of FRTs, which in turn means that 80% could mean very different inferences for different technologies.

## Transparency and Privacy

Transparency has been found to be a major issue in the implementation of FRTs. While this might not be a primarily technical/design issue, it still has severe ramifications and must be discussed. The usage of FRTs in law enforcement has largely been opaque. From the choice of the arbitrary threshold number, the distribution of CCTV cameras to conduct surveillance to the procurement of the FRTs from private players, there have been numerous implementation choices that lean towards a sense of opaqueness with respect to decision making. A report had shown that the distribution of CCTV cameras to conduct surveillance was not uniform[36]. The paper noted, that while one couldn't conclude if the non-uniformity was intentional, it definitely caused disproportionate surveillance of communities that lived in the densely surveilled area. Another paper noted that the role of private players aiding procurement of FRTs points out the possibility of loss of transparency[37]. Since there exists no standard to adhere to and no regulatory body, it becomes a mammoth task to ensure that these procurements work in the best interest of the general public. While RTIs have been giving us insights into the implementation of FRTs, the lack of audit reports on these procurements highlight the possibility of there being no scope for any public scrutiny for a downstream task that affects the general public in a massive way.

With these applications there is an overarching privacy risk. The development of FRT algorithms requires access to large datasets of pictures, videos, or any other graphic corpuses. The question arises regarding how private corporations have accessed such data in India. Furthermore, there are also concerns around whether the private sector should have unbridled access to vital

---

[35] ibid, The responses to the RTIs filed by the Internet Freedom Foundation can be found here; <https://drive.google.com/file/d/1EeF-d5Z1pZVp6SZ5S-53iPZKNqen60Zj/view?ref=static.internetfreedom.in>.
[36] ibid.
[37] ibid.

Centre for Responsible AI

biometric data (including facial scans) of individuals to design such technology purportedly for state security purposes, in complete opacity.[38] However, even if it is being used entirely by the public sector, issues of privacy remain. As per the law declared by the Supreme Court, the state (which includes the central, state, local governments as well as their associated agencies) are liable to conduct a proportionality analysis before the right to privacy can be infringed upon. As part of this analysis, the infringement must be validated by an enacted legislation. Secondly, the infringement must be for a legitimate aim. Thirdly, the extent of interference must be proportionate to the need for such interference. Finally, there must be procedural guarantees against the abuse of such an interference. Establishing proportionality of the infringing measure, which will include establishing efficacy of this method (that is, FRT) over other methods which were previously being used for the upkeep of law and order, is upon the state.[39] As they are used currently, it becomes important to see if these solutions comply with the three-pronged test laid out by the Supreme Court in the *K.S. Puttasamy* judgement[40] or with newer legislation like the Digital Personal Data Protection Act, 2023[41].

## *Operationalising PAI-FRT*

The identification of stakeholders is an integral part of operationalising PAI. The stakeholders help us identify important relations in different decision contexts and decide the order of importance. This is incredibly important for the framework proposed in Paper 1 as involving the relevant parties for taking robust decisions takes centre stage. With the given context of Law Enforcement let us identify a few stakeholders based on the scholarship surveyed in Paper 1 to ground our framework. While stakeholder identification is a subjective process, some common stakeholders identified usually include patients, clinicians, managers, executives, clinical assistants and payers.

---

[38] ibid.
[39] While the infringement of a right has to be established by the petitioner, the justification of that infringement has to be established by the state. On the burden of proof generally, see Saghir Ahmad vs The State Of U. P. And Ors.; 1954 AIR 728, 1955 SCR 707 & Saghir Ahmad vs State on 18 October 1960; AIR 1961 ALL 507.
[40] *Justice K.S. Puttaswamy (Retd.) & Anr. vs. Union of India & Ors.*; (2017) 10 SCC 1.
[41] The Digital Personal Data Protection Act, 2023, <https://www.meity.gov.in/writereaddata/files/Digital%20Personal%20Data%20Protection%20Act%202023.pdf>.

Paper 1 talks about using relations of urgency, legitimacy, power and harm to identify the stakeholders. We can use these factors to identify stakeholders for our example, as listed below:

- Law enforcement - relations of power and legitimacy as they get to use the technology on the general public and terms of implementation. This can also include lawmakers who are authorised and required to place guardrails or integrate participation in policies.
- Civil Society - relations of legitimacy as they have the right to critique policy implicitly enshrined in the Constitution and relations of harm as they are directly affected by said policy
- Private FRT Companies - relations of power as they are the source of the technology
- Undertrials - relations of legitimacy, urgency and harm as the results of FRTs used against them gives them ground for a legitimate claim about the accuracy, an urgent claim as they are under trial and a claim against harm as these results could result in an unfavourable judgement for them
- Judiciary - relations of power as they get to decide the validity of the results of FRTs on a case by case basis and relations of legitimacy as they are given the power to make judicial decisions by the Constitution of India

As proposed in Paper 1, PAI can be operationalised by implementing a Decision Sieve, spreading various aspects of our solution across two planes. The horizontal plane would contain the stakeholders, and provide the base for the sieve. The vertical plane would host the flow of information and decisions from one phase to the next.

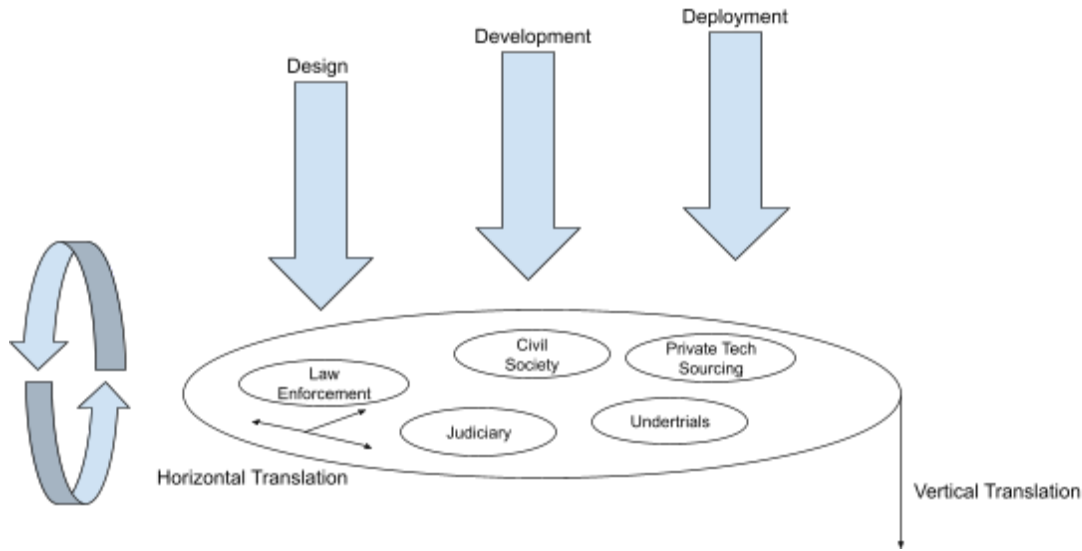

We run through the relevant stakeholders every pass. Each phase (Design, development, deployment) may contain multiple passes. For each pass, the relevant stakeholders share information amongst each other. As each stakeholder occupies their own niche, the framework recommends an aspect of horizontal translation to ensure that points of significance are conveyed and are contextualised. This is the first step towards ensuring equity in the decision making process. Each pass generates decisions and actionable information that are sequentially passed on to the next pass/phase. This information needs to be translated as per the requirements of the next pass, ensuring consistency and clarity, helping the next pass function efficiently. The vertical and horizontal translations are integral to the framework as they contextualise the information and promote participation.

Each phase can be viewed as multiple sequential passes that help iron out decisions and give the solution an overarching structure. The planes try to ensure that the participation of a stakeholder isn't limited to any one phase and can be consulted for decisions. Each phase requires a certain set of decisions to be made with respect to the solution. The Decision Sieve acknowledges that some decisions' span could even be formed over multiple phases. These compound decisions are made by using the Sieve recursively to answer the minutiae and build up to the more complex questions that would be easier to answer thanks to aggregation of smaller decisions capturing the nuance of the decision. The aim of the framework is to ensure these decisions are made and propagated in a meaningful manner and that the aspirations of all involved groups are well represented.

For example, in the case of FRTs, we can take a look at the different decisions one must take across the different phases of AI development. To assess whether FRT should be used, lawmakers, civil society, and representatives from affected population(s) must work with why policing warrants (or not) such use of FRTs. Following which, to make the decision of what data the model should be trained on, it would be vital to involve Civil Society, Law Enforcement and the Facial Recognition Technology provider. While the class distribution and procurement of data are primarily overseen by the Tech Provider and Law Enforcement respectively, the inherent constraints with respect to those processes can be translated to all three parties so that all parties find their interests well represented and take an informed decision. Similarly, implementation in the form of CCTV surveillance and its distribution could also be a decision that is made through the participation of relevant stakeholders. While some decisions are limited to a single phase, there are some questions that evolve across phases that might at different points need the opinion of different stakeholders. The iterative process accounts for such shifts, and helps collate all these decisions without divorcing any relevant stakeholders from the decision making process. Questions like what counts as a positive match can now be answered iteratively, through multiple passes of horizontal and vertical translation to form a composite decision that reflects every stakeholder's views.

While the framework sets the tone for incorporating participation, the way decisions are made would vary across contexts and organisations. While Civil Society's opinion might be given more weightage than the Tech Provider's when it comes to surveillance and privacy, the judiciary and law enforcement's opinion might hold more weight in the decision regarding feasibility or the validity of FRT matches as evidence. Contexts and pre-existing hierarchies direct decision collation and the subsequent transfer of information. The framework ensures that each different viewpoint is represented and contributes to the decision making process.

The issues associated with the usage of FRTs are now mitigated through Participatory Design. The decisions made are representative of all relevant views, inclusive, interpretable and transparent. This helps mitigate issues such as bias and fairness through equitable representation of diverse views, and transparency and interpretability through a clear log of which combination of choices contributed to the final decision. It also reduces friction between outcomes and users as the users have a direct hand in the development of the outcomes. This also serves to aid Grievance Redressal as there is increased clarity as to which decision and choices contributed to the grievance at hand. The decisions also help set the standards for monitoring as the decisions retain the nuances of the various views that were exhibited in discussion and would serve as north stars that can be used to establish standards for acceptable performance. Thus, a participatory approach to AI governance could form the foundation of adopting and achieving Responsible AI implementations for various solutions.

## *What does this mean in the Indian Context?*

At the time of writing, the only official discourse on FRTs has come from NITI Aayog's discussion paper titled "Responsible AI for All" that outlines principles for responsible artificial intelligence (AI) adoption in India. It takes a use case approach and uses FRTs to ground their framework. NITI Aayog also highlights certain key principles that form the basis of Responsible AI. The key principles include:

- Inclusivity and Non-Discrimination: Advocating for AI systems that promote inclusive development, economic growth, and social progress to benefit all segments of society.
- Equality: Focusing on ensuring that the benefits of AI are accessible to all citizens, bridging the digital divide, and preventing any form of discrimination.
- Protection and Reinforcement of Positive Human Values: Encouraging the use of AI for creating public value, addressing societal challenges, and promoting the well-being of citizens while emphasising the importance of using AI to enhance individual and collective human capabilities rather than substituting or diminishing them.
- Privacy and Security: Highlighting the significance of protecting privacy rights and ensuring that individuals have control over their data.
- Transparency and Accountability: Promoting the development and deployment of AI systems that adhere to ethical standards, are transparent, and can withstand adversarial attacks.

In the use-case based approach the discussion paper goes over the applications, issues and possible questions that FRTs give rise to. Their approach to parse through these insights is dictated by the principles elucidated earlier. Taking these principles as a north star to establish a sense of direction, we can assert that PAI as a framework aligns with the principles of what constitutes Responsible AI in the Indian context. It promotes agency, empowering all stakeholders to put their views forth, ensuring that the value added has a positive equitable social impact. Furthermore the principles put forth by NITI Aayog would help steer the decision made as a result of PAI discussions in the right direction.

## *LLMs in Healthcare*

With the growing uptake and integration of AI into different domains, the Healthcare sector has been actively looking into Large Language Models ('LLMs') to take the burden of clerical and largely trivial tasks that involve very little skill. From tightening the financial aspects of a Hospital through ICD code prediction and automated billing to prevent revenue leakage, to Diagnoses aid and predictors to reduce the load on doctors, AI and specifically LLMs are quite sought after as a solution. LLMs are complex deep learning models that are trained on massive amounts of textual data through self-supervised learning to learn statistical associations and generate text tokens based on inputs given to them. Recent advancements in the field of Natural Language Processing such as GPT-4 (ChatGPT), Google PaLM, Google Gemini (Bard) and Meta's Llama 1 & 2 pushed the boundaries of performances in text processing tasks and conversational agents. Several application sectors are now making use of these LLMs for research as well as commercial applications by zero-shot manner or fine tuning the LLMs.

While generic LLMs have attained significant improvement in performances in natural language tasks, domain-specific LLMs have been on the rise for developing dedicated functionality in a particular sector. For instance, Biomedical LMs are models trained on biomedical literature or textual data that have potential applications in the field of biomedicine and healthcare as these models understand the domain-specific jargons and context. LM variants of BERT like BioBERT, PubMedBERT, BioLinkBERT, BioGPT, etc have been used in the past to showcase its potential in the field of biomedicine for various tasks such as Question Answering, Named Entity Recognition, Document Classification, etc. Google's MedPaLM2 is a recent variant of PaLM model fine tuned on biomedical textual data for performing natural language tasks in the biomedical domain and has shown its potential in various benchmark tasks based on multiple choice questions, question answer generation and language understanding. Recent publications have highlighted potential applications of LLMs in medicine and healthcare and have categorised them into Clinical Applications, Administration Applications, Research Applications and Educational Applications.

## *How LLMS Work*

Large Language Models (LLMs), such as GPT-3, operate by leveraging deep neural networks to understand and generate human-like text. These models consist of millions or even billions of parameters, enabling them to capture complex patterns and relationships in language. During training, LLMs learn from vast datasets, adapting their parameters to mathematically predict the next word in a sequence[42]. This process enables the model to grasp grammar, context, and semantic nuances[43]. When given a prompt, the LLM utilises its learned knowledge to generate coherent and contextually relevant text, showcasing its ability to perform various language tasks, from translation to creative writing. Despite their impressive capabilities, these models are not infallible and may exhibit biases or generate inaccurate information (hallucinations), emphasising the importance of careful use and ethical considerations[44].

## *Scope of LLMs in Healthcare*

In this paper, we consider healthcare applications that can be beneficial for both Clinical as well as Administrative applications in a Healthcare setting.

### *Patient Summary generation*

Patient summaries offer concise overviews of medical history, diagnoses, treatments, and other pertinent information, serving as crucial information reference documents for clinicians, patients, and even researchers. There could be various kinds of summaries such as discharge summaries, nursing summaries, doctor's summaries and so on. These summaries usually are generated based on all the clinical and demographic information that are present as part of Electronic Health

---

[42] Brown, Tom, et al. "Language models are few-shot learners." *Advances in neural information processing systems* 33 (2020): 1877-1901. <https://arxiv.org/abs/2005.14165>.
[43] Vaswani, Ashish, et al. "Attention is all you need." *Advances in neural information processing systems* 30 (2017). <https://arxiv.org/abs/1706.03762>.
[44] Bender, Emily & Gebru, Timnit & McMillan-Major, Angelina & Shmitchell, Shmargaret. (2021). On the Dangers of Stochastic Parrots: Can Language Models Be Too Big?. 610-623. 10.1145/3442188.3445922.

Records (EHRs) of patients. Traditionally, these summaries are manually generated by physicians or other health personnel and is quite a time-consuming, mechanical, and subjective process prone to differences in writing styles, templates and at times include some inconsistencies and rare omissions of critical information/condition. In this regard, there is huge potential for LLMs to help the healthcare system by generating effective summaries from electronic health records of patients. Such models can be trained on vast collections of structured/unstructured EHRs, unstructured clinical notes, and even biomedical research papers, enabling them to learn the concepts and complexities of medical language and the essential elements of comprehensive patient summaries. These models can then be tasked with generating summaries automatically, extracting and presenting relevant information based on the patient data within the EHRs. While these summaries are useful for caregivers such as doctors and nurses for better streamlining their clinical practice and care provision, they also have great use for the hospital administration tasks such as medical records maintenance, insurance claims, pharmacy and other inventory orders, appointments and follow-up.

### *Disease diagnosis and ICD coding (Billing and insurance claims)*

AI models for developing disease diagnostic models are one of the most popular healthcare tasks being investigated by researchers. Several existing approaches have explored using Deep Learning architectures ranging from basic MultiLayer Perceptrons to Transformers based Language Models for building disease diagnosis models[45]. While some models have been proposed to specifically target diseases, other models have attempted to predict the International Classification of Diseases (ICD) groups or codes for the patients based on the clinical data available as part of the EHRs. LLMs have the potential to process and learn from large structured/unstructured EHRs to mine patient-specific patterns in conditions, symptoms and other clinical information to provide the physicians with disease diagnosis possibilities, enabling personalised medicine and even rare disease diagnosis. LLMs capture this by mining symptom-disease relationships and largely aid healthcare professionals in narrowing down possibilities. AI based Personalised Medicine could revolutionise the healthcare sector as it can provide the doctors with diagnostic possibilities and other recommendations based on patient-specific characteristics. By considering factors like genetics, lifestyle, and medical history (EHRs), LLMs can estimate a patient's personalised risk of developing specific diseases, enabling the healthcare personnel to take proactive measures and preventative strategies. Based on patient-specific risk profiles, LLMs can even potentially recommend personalised diagnostic tests or strategies, optimising the diagnostic process for each patient. While these models are discussed in terms of the diagnosis process, similar models can be used in the administrative sections as well for applications like ICD coding for medical records and insurance claims. Deciding the ICD code is crucial for tasks like insurance claim decisions as ICD codes can not only describe the diagnosis/condition, but it can convey the severity of the condition as well. While insurance claim amounts are decided based on severity of conditions, ICD coding becomes an important task for the patients as well as the insurance providers in terms of the financial aspects. Currently a manual task, ICD coders often make mistakes and there are huge losses caused to either parties due to this. Therefore, the accuracy of the ICD code prediction tasks become extremely crucial and this is where LLMs' effective abilities in capturing statistical correlations between text tokens and semantic natural language understanding, provide us with great potential to solve this task.

---

[45] *Note: Various approaches employing Deep Learning architectures have been explored for disease diagnosis. Deep Learning involves using neural network structures capable of learning complex patterns from data. Two commonly utilised architectures are MultiLayer Perceptrons (MLPs) and Transformers. MLPs consist of multiple layers of interconnected nodes (artificial neurons) and are effective at learning hierarchical representations of data. Transformers, on the other hand, utilise attention mechanisms to weigh the significance of different parts of the input data and have shown significant success in natural language processing tasks. These architectures are trained on large datasets of medical information to develop models capable of accurately diagnosing diseases.*

### *Advisory Chatbots*

LLMs, with their remarkable ability to learn from huge amounts of various kinds of text data such as patient records or EHRs, biomedical research literature/articles, etc. paired with their abilities to process and generate human-like text, offer immense potential in revolutionising how patients interact with healthcare systems and manage their health routines and procedures. The medical chatbots can interact with patients and can potentially streamline a number of tasks within a hospital ecosystem – both in terms of clinical as well as administrative tasks. Chatbots built on top of effective architectures like LLMs can not only potentially identify diagnoses possibilities, but also can provide first aid advice, facilitate specialist appointment scheduling or even alert emergency services at times of need. Unlike traditional symptom checkers, LLMs can engage in dynamic, natural language conversations, tailoring questions based on individual responses and medical history. This can give a more personalised touch to the interacting patient and also enable personalised medicine approach which can ensure more accurate symptom assessment and potential diagnosis suggestions, empowering patients to seek appropriate care with specialists. It also has the potential to continuously monitor the patient with medication reminders and feedback mechanism, scheduling follow up appointments and even help with mental health of the patient by providing calming advice to help at times of anxiety. LLMs can potentially even remove language barriers as they can be trained and deployed in multiple languages empowering the chatbot so that support can be provided in the patient's language.

## Issues:

### High Cost of failure

Though there have been metrics to evaluate performance of AI and Language models such as accuracy and precision, the model when deployed in the real world will affect users/stakeholders differently as per the sector in which the AI application was deployed. Thus, the need for metrics to measure failure and its social impact arises. It is an integral part of contextualising the real world impact of any deployment. In the case of healthcare operations which are directly linked to the patient's care, there is a high cost of failure. Though these operations have varying degrees of impact on a patient's wellbeing, with the cost of failure directly correlated, the impact of a mistake made by an AI based solution is very significant in a lot of use cases. For that reason we must limit the scope of our use cases to tasks/operations that:

- in the case of applications that have a direct effect (disease diagnosis or prediction), operate with AI-in-the-loop. AI-in-the-loop emphasises that AI should augment humans, but humans should always remain at the centre of decision-making. That is to say, the decisions or predictions made by the AI solution are used under the oversight and discretion of domain experts, in this case doctors and nurses.
- are administrative or clerical in nature. These tasks are non-clinical, thus not impacting the patient care process. This would help improve the quality experience for the patient's without directly affecting their treatment.
- are clinical tasks that can be vetted and cleared and thus do not directly impact patient care. This includes operations like generating discharge summaries etc

By understanding and acknowledging the risks and limitations of these applications we would be able to better implement these solutions and safeguard stakeholders.

## LLM Hallucinations:

LLMs have been found to hallucinate facts, information and data. There are emerging techniques like Retrieval Augmented Generation (RAG) that allow for hallucinations to be minimised, but an LLM is unlikely to be 100 percent rid of hallucinations. The trust in AI based solutions is based on empirical analysis, backed by data. The possibility of hallucinations in applications that require factual and accurate information makes it an incredibly risky undertaking. This can have significant implications, particularly in healthcare applications, where accurate information is crucial. For example, a medical chatbot hallucinating symptoms or providing inaccurate treatment suggestions may pose serious risks to patients[46]. Ethical concerns arise, emphasising the need for rigorous validation and careful deployment of LLMs in critical domains, as their hallucinatory outputs could impact decision-making and patient outcomes[47]. Ensuring transparency and accountability in LLMs is essential to mitigate these potential pitfalls.

## Fairness and Bias:

As noted in the previous paper, biases are not uncommon in ML/AI based solutions. Data biases and algorithmic biases are often found in ML systems.[48] Class imbalances and algorithmic preferences can lead to disproportionate outcomes in downstream fields. For a domain like Healthcare, the possibility of non-equitable patient welfare is incredibly concerning[49]. It goes against the principles of medical ethics and must be mitigated. Though debiasing is an option, usage of the current crop of LLMs should be done with complete understanding of what the LLMs limitations are. Having said that, generating summaries give high quality prompts, whose results are subject to human supervision, would tap into the powerful capabilities of LLMs without putting patients at risk.

---

[46] McCoy, T. H., Castro, V. M., Cagan, A., Roberson, A. M., Kohane, I. S., & Perlis, R. H. (2018). "Sentiment measured in hospital discharge notes is associated with readmission and mortality risk: an electronic health record study." PloS One, 13(11), e0206843.

[47] Bender, E. M., & Friedman, B. (2018). "Data statements for natural language processing: Toward mitigating system bias and enabling better science." Transactions of the Association for Computational Linguistics, 6, 587-604.

[48] Nima Kordzadeh & Maryam Ghasemaghaei (2021): Algorithmic bias: review, synthesis, and future research directions, European Journal of Information Systems, DOI: 10.1080/0960085X.2021.1927212.

[49] Riley, Wayne J. "Health disparities: gaps in access, quality and affordability of medical care." *Transactions of the American Clinical and Climatological Association* vol. 123 (2012): 167-72; discussion 172-4.

### Opacity of AI Decision Making:

Explainable AI is a fundamental aspect of Responsible AI.[50] [51] Explanations make AI solutions easier to understand and ground them in their respective domains. While robust and accurate predictions are integral to an AI solution, explainability of said solution helps in understanding the choices of the model in the terms of the domain. What this means is, explainability gives us greater insight into the choices of the model and helps us review the working of it in a much more efficient manner.

### AI in the loop v/s Human in the loop:

With the large-scale adoption of AI in healthcare, the value of robust standard operating procedures being put in place is highlighted. AI is a powerful tool with which one can significantly optimise tasks. However, neither is the solution one size fits all nor is it a balm that solves issues. It too has limitations, some of which can have significant repercussions. To offset these issues, workflows that operate around the capabilities and limitations of the AI solution add great value. One such proposed framework/workflow is AI-in-the-loop. It is a slightly altered perspective on Human-in-the-loop. Human-in-the-loop is a blend of supervised machine learning and active learning where humans are involved in both the training and testing stages of building an algorithm. While the decision making is largely centred around machines with importance given to human feedback, AI-in-the-loop emphasises that AI should augment humans, but humans should always remain at the centre of decision-making. This allows us to retain some sense of regulation on the quality of outputs of the AI solution and also allows for a more seamless integration into different domains, which is something human-in-the-loop might struggle to achieve.

---

[50] Baker, Stephanie, and Wei Xiang. "Explainable AI is Responsible AI: How Explainability Creates Trustworthy and Socially Responsible Artificial Intelligence." arXiv preprint arXiv:2312.01555 (2023).
[51] ibid.

## *Operationalising PAI- LLM*

Borrowing from the same points used in 'Operationalising PAI- FRT', the following stakeholders emerge:

- Doctors - relations of legitimacy and power as domain experts and clinical decision makers respectively.
- Patients - relations of urgency as they are in need of treatment and relations of harm as adverse decisions could harm them.
- Funders - relations of power as they are the principal financiers.
- Legal Team - relations of legitimacy and power, as they are experts whose legal opinions are highly weighted.
- Developers - relations of legitimacy as the experts in their domains.
- Administration - relations of power and relations of harm as the new technology could adversely affect their work.

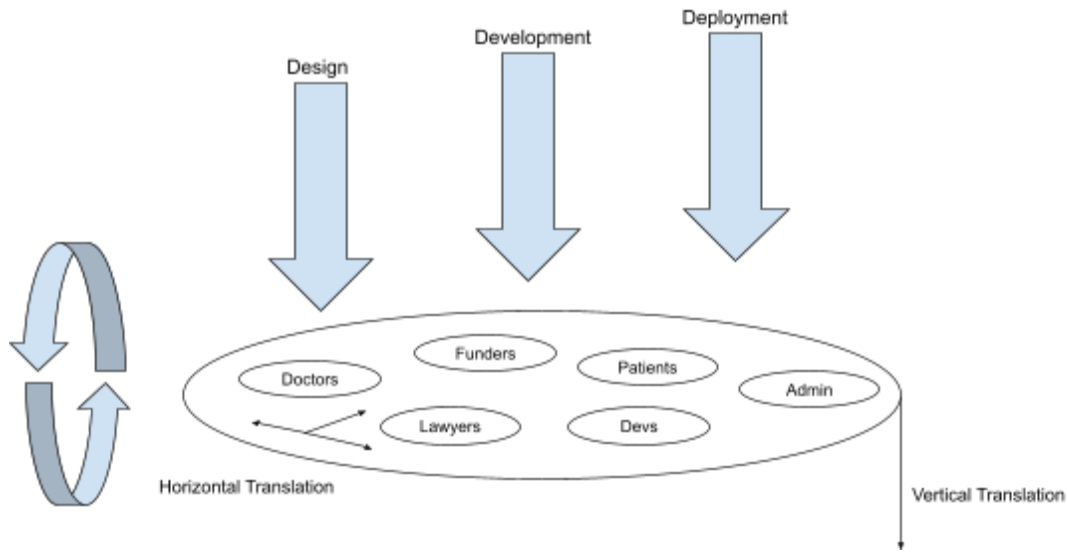

Borrowing from use case 1, the same steps will be used to operationalise the Decision Sieve and implement its core tenets for all the phases of creating an AI based solution.

For example, if one was to use LLMs to generate patient discharge summaries, one could involve patients, doctors and the developers to make decisions on the feasibility and integration of the LLM that would generate these summaries. While the feasibility is primarily governed by the doctors (in terms of the domain feasibility) and developers (in terms of technological feasibilities, the constraints can be translated to all three parties so as to take an equitable and well informed decision. Primarily a Design question, the responses to the sub-questions that are part of the decision on feasibility might evolve as the process progresses and require different opinions, for example the funders. The iterative process accounts for such shifts, and helps collate all these decisions without divorcing any relevant stakeholders from the decision making process.

While the framework sets the tone for incorporating participation, the way decisions are made would vary across contexts and organisations. While the doctor's opinion might be given more weightage than a developer's when it comes to the medical implication of a solution, the lawyer's opinion might hold more weight in the decision regarding legal feasibility or the possibility of repercussions of using LLMs. Contexts and internal hierarchies direct decision collation and the subsequent transfer of information. The framework ensures that each different viewpoint is represented and contributes to the decision making process.

Through this manner of participation, we can now mitigate a lot of the aforementioned risks. Participation provides for a space where bias can be mitigated through representation. And since all views and constraints have been discussed, the volume of grievances can also be expected to decrease. Issues like poor design and implementation challenges are dealt with through comprehensive discussions and decision collation. A clear decision log can help account for which combination of choices contributed to a decision, making decisions incredibly interpretable. And while PAI attempts to reduce grievances, the grievances that might occur now can be resolved in a more efficient manner due to interpretable decision making and accountability. This would also serve to aid the integration into pre-existing forms of grievance redressal.

## *What does this mean in the Indian Context?*

The Indian Council for Medical Research (ICMR) has released the Ethical Guidelines for AI in Biomedical Research and Healthcare[52]. Since AI cannot be held accountable for its decisions, an ethically sound policy framework is essential to guide AI technology development and its application in healthcare. The ICMR guiding document stated that as AI technologies develop and are applied in clinical decision-making, it is important to have processes that discuss accountability in case of errors for safeguarding and protection. The document tries to tackle concerns about potential biases, data handling, interpretation, autonomy, risk minimisation, professional competence, data sharing, and confidentiality.

The document highlighted ten key patient-centric ethical principles for AI applications. These principles include accountability and liability, autonomy, data privacy, collaboration, risk minimisation and safety, accessibility and equity, data quality optimisation, non-discrimination and fairness, validity and trustworthiness. Informed consent and governance of AI tools in thebrhealth sector are other critical areas highlighted in the guidelines.

Ensuring these principles are observed and implemented can be achieved through PAI. Using these principles as a north star could be highly beneficial to the process of decision collation. Through active participation of relevant stakeholders and well documented participation, issues like accountability and liability would be much easier to tackle from a governance perspective. PAI would also be greatly beneficial in ensuring the AI solution adheres to principles like accessibility, equity, non-discrimination and fairness, trustworthiness etc. Through active participation we can ensure the resultant AI solution represents the perspectives of everyone involved and can have an equitable and positive effect in a domain like healthcare where working with the patients is a core tenet.

---

[52] ICMR, Ethical guidelines for application of Artificial Intelligence in Biomedical Research and Healthcare, 2023, 978-93-5811-343-3.

## *Conclusion*

This paper may be summed up in the following propositions- *firstly,* that AI offers various benefits by way of efficiency; *secondly,* that reaping these benefits depends upon how algorithms are implemented; and *thirdly,* that a participatory approach can help reap these benefits better. These propositions were derived partly from Paper 1, and tested in this paper through the FRT and LLM examples. The demonstration has been more conceptual in nature, rather than empirical. Future research may build upon this and test the decision sieve in actual settings in either law enforcement or healthcare.

The sieve is illustrative and simplified. Full implementation will, in all probability, result in more complexities than has been portrayed here. Despite this, the decision sieve is conceptually a scalable model (with minor changes as considered necessary), and can be used in larger settings. This is because the interests affected, the benefits to be attained by deployment of AI (efficiency), the costs to be borne by such deployment, and the possibility of these costs being absorbed/reduced via a participatory approach will remain essentially unchanged. As long as this condition is satisfied, the decision sieve affords a systematic method of planning and implementing a participatory approach in AI development and deployment.

A second caution is that this paper has proceeded on the assumption that AI (especially FRT), complimented with appropriate participatory input, will solve all issues. This is clearly not the case. Even if implemented well, these systems can still cause harm- say, in the form of misidentification in an individual case, and the collective harm suffered from total surveillance. A participatory approach does not take away from the need to inquire into whether an AI solution is necessary in the first place. It only shapes the contours of, if this question is answered in the positive, the method by which design and deployment will take place.

For any queries relating to this paper, please reach out to ambreesh@cerai.in

www.vidhilegalpolicy.in

www.cerai.iitm.ac.in

Vidhi Centre for Legal Policy
A-232, Defence Colony
New Delhi – 110024

Robert Bosch Centre for Data Science and AI, 5th floor, Block II, Bhupat and Jyoti Mehta School of Biosciences, Indian Institute of Technology Madras, Chennai-600036

011-43102767 / 43831699

+914422574370

Centre for Responsible AI